\documentclass[aps,prl,twocolumn,superscriptaddress,nofootinbib,
nobalancelastpage,showpacs,nobibnotes]{revtex4}
\usepackage{epsfig}

\begin{document}
\title
{Instabilities in neutrino systems induced by interactions with scalars
}

\author{R. F. Sawyer}
\affiliation{Department of Physics, University of California at
Santa Barbara, Santa Barbara, California 93106}

\begin{abstract}
If there are scalar particles of small or moderate mass coupled very weakly to Dirac neutrinos, in a minimal way,
then neutrino-anti-neutrino clouds of sufficient number density can experience an instability in which
helicities are suddenly reversed. The predicted collective evolution is
many orders of magnitude faster than given by cross-section calculations.
The instabilities are the analogue of the ``flavor-angle" instabilities (enabled by the Z exchange force) that may drive very rapid flavor exchange among
the neutrinos that emerge from a supernova. These exchanges do require a tiny seed in addition to the scalar
couplings, but the transition time is proportional to the negative of the logarithm of the seed strength, so that
the size of this parameter is comparatively unimportant. For our actual estimates we use a tiny non-conservation
of leptons; an alternative would be a neutrino magnetic moment in a small magnetic field. The possibility
of a quantum fluctuation as a seed is also discussed.
Operating in the mode of putting limits on the coupling constant of the scalar field, for the most minimal
coupling scheme, with independent couplings to all three $\nu$, we find a rough limit on the dimensionless coupling constant for a neutrino-flavor independent
coupling of $G<10^{-10}$, to avoid the effective number of light neutrinos in the early universe being essentially six.
If, on the other hand, we wish to fine-tune the model to give a more modest excess (over three) in the effective 
neutrino number,
as may be needed according to recent WMAP analyses, it is easy to do so.


\end{abstract}
\maketitle

\section{1. Introduction}
In the last few years there has been an accumulating literature focused on 
the possibility that the neutral current $\nu-\nu$ interaction 
can induce rapid collective flavor exchanges of physical
interest in situations in which there are initial flavor-energy correlations or
flavor-angle correlations \cite{ful1}-\cite{rfs}. Up to now the application has been to
the supernova neutrino pulse, where, apart from the dream that a galactic
event will bring us information on neutrino properties directly from
terrestrial detection, the flavor-energy correlations
may matter to the explosion dynamics or to the possible R-process nucleosynthesis 
on the periphery of the supernova. 

The purpose of the present paper is to show that the same kind of effect can enter in models in which neutrinos interact with scalar particles. Such interactions have been the subject of much speculation \cite{georgi}-\cite{hann2}.
We believe that the exploration of different models of such an interaction
will lead to a whole zoo of exotic phenomena, but we focus on the simplest
example in this preliminary study, and we shall consider only one flavor of Dirac neutrino
coupled to a single scalar field. The effects that we find are of most interest with respect to
the early universe, in the period immediately prior to
light element nucleosynthesis. In this system we have  
a thermal distribution of left-handed neutrinos, and their anti-particles, in nearly equal
number.
In standard theory a negligible number of right handed light neutrinos 
have been produced up to that early time, through a combination of collisions and the tiny neutrino
mass term.

The collective effect that can come into play through the medium of the scalar particle exchange is the sudden and
simultaneous creation of right-handed $\nu$'s and left handed $\bar \nu$'s, keeping the lepton number equal to zero,
and with a rate orders of magnitude greater than one would predict from a perturbative cross-section.
The constraint on the model would thus be the need to avoid excessive production
of right-handed $\nu's$ and their antiparticles prior to neutrino decoupling in the big bang; that is, to avoid
increase of the effective number of neutrino species to above the limits set by
cosmology. In this system the collective effects of the $\nu-\nu$ force arising from $Z_0$ exchange vanish,
owing to the cancellation between the particle and anti-particle terms in the effective potential.   

As in the supernova case, the surrounding medium in our early-universe environment is filled with 
other particles than neutrinos. We shall consider only cases in which collision times for the neutrinos to scatter on these other
particles are long compared to the instability times. Then the only
effect of the other particles is through the constant background neutrino potential 
coming from from the neutral current interactions with electrons and positrons (the interactions with nucleons
being flavor independent and therefore irrelevant). In the early universe the number of electrons so nearly
equals the number of positrons that this effect is negligible.

We first take a rather long detour through the existing lore of 
collective instabilities that derive from the $Z_0$ force. It is not our intention
to write a review article, but we do need to discuss in detail what is the same and what is different
in the scalar-mediated case. Furthermore, we wish to underscore two areas in which one might question the assumptions of the
whole enterprise, including the work in refs. \cite{ful1}-\cite{rfs}: 1) the replacement of the
interaction Hamiltonian by a ``forward" Hamiltonian; 2) after this replacement, and the extraction of
equations of motion, a ``mean field" assumption. In the case of potential issue \#1, the issue is, in a pragmatic way,
de-fused (but not put to rest) by staying in domains in which the interesting physics takes place over distances
much less than a mean-free path as calculated from cross-sections. In the case of the mean-field
assumption, we address the question directly from simulations with a finite number of neutrinos in a box.
Assemblages of a few hundred neutrinos are computationally accessible on a desktop, and with this number
we can support the mean-field method in some of the simpler coupling schemes. We return to these
issues in section 4.
\section{2. Forward Hamiltonian from $Z_0$ exchange.}
In all cases we begin with an effective interaction Hamiltonian, $H_I$, arising from an exchange
of a boson. 
Then if we are operating in , say, the MeV region of $\nu$ energies the $\nu-\nu$
interaction mediated by $Z_0$ exchange can be taken as a local four-Fermi
interaction. In the case of our scalar exchanges we do want to cover the possibility
of exchanged masses in the region below  1MeV and thus the propagator for
the scalar particle contributes momentum dependence.

The key to all of the literature cited in refs.\cite{ful1}-\cite{rfs}
is the replacement of $H_I$ by a ``forward" Hamiltonian. In this reduction we expand
each field in plane wave modes; each term now is characterized by a momentum
sequence, ${\bf p_1,p_2\rightarrow p_3, p_4}$. The forward interaction Hamiltonian
is defined by omitting from the sum all terms except those in which ${\bf p_1=p_3; p_2=p_4}$
or  ${\bf p_1=p_4; p_2=p_3 }$. Then in discussing the evolution of some initial state
under the influence of $H_I$ we can ignore the kinetic term altogether. The only physical effects
that ensue are now the exchange of flavors among momentum states (or we could equally say:
the exchange of momenta among flavor states) in the case of the $Z_0$ model, or the alteration of
helicities, with momenta unchanged, for the case of the scalar interactions. In dynamical calculations
the free Hamiltonian $H_0$, is now irrelevant; we deal entirely in a subspace of states of equal unperturbed
energy. 

These ``forward" interaction Hamiltonians can now be completely expressed in terms of bilinear forms
in neutrino creation or annihilation operators, $a^\dagger_i ({\bf p})\,a_j ({\bf p})$, where $\{i,j\}$
indices are of  flavor for the $Z_0$ case, and of helicity for the scalar case,
\begin{eqnarray}
\rho_{i,j}({\bf p})=a_i({\bf p})^\dagger \, a_j({\bf p})
\nonumber\\
\bar \rho_{i,j}({\bf p})= \bar a_j({\bf p})^\dagger \,\bar a_i({\bf p}) \, ,
\end{eqnarray}
with the commutation relations,
\begin{eqnarray}
[\rho_{i,j}({\bf p}),\rho_{k,l}({\bf p'})]=[\delta_{i,l}\rho_{k,j}({\bf p})-\delta_{j,k}\rho_{i,l}({\bf p})]\delta_{\bf p,p'}\,,
\nonumber\\
\,
\nonumber\\
\,[\bar \rho_{i,j}({\bf p}),\bar \rho_{k,l}({\bf p}')]=[-\delta_{i,l}\bar\rho_{k,j}({\bf p})+\delta_{j,k}\bar\rho_{i,l}({\bf p})]\delta_{\bf p,p'}\,.
\label{com}
\end{eqnarray}
In terms of these operators the forward Hamiltonian is,

\begin{eqnarray}
&H_{\nu \nu}(\rho)= {\sqrt{2}G_F\over V }\sum_{{\bf p, q} }~\sum _{i,j}
 [1-\cos (\theta_{\bf p,q})]
\nonumber\\
&\times \Bigr[ \Bigr(\rho_{i,j}({\bf p})-\bar \rho_{i,j}({\bf p})\Bigr )
\Bigr(\rho_{j,i}({\bf q})-\bar \rho_{j,i}({\bf q})\Bigr )~~~~~~~~~~
\nonumber\\
&~~~~~+\Bigr(\rho_{i,i}({\bf p})-\bar  \rho_{i,i}({\bf p})\Bigr )
\Bigr(\rho_{j,j}({\bf q})-\bar \rho_{j,j}({\bf q})\Bigr )\Bigr ]\, .
~~~~~~~~~~~~~~~~
\label{ham}
\end{eqnarray}
where the indices $i,j$ range over the values $e,\mu,\tau$.

We divide the momentum space into $N_B$ bins, each of them containing $N_\nu/N_B$ neutrinos, and distinguished
by the superscript $(\alpha)$.
With few exceptions the instability literature is based on two-flavor models
$i=\{1,2\}$. In this case it is convenient to use angular momentum-like collective variables $S^{(\alpha)}_0, S^{(\alpha)}_\pm, 
S^{(\alpha)}_3$, formed from the $\rho({\bf p})_{i,j}$ by summing over the region of momentum space, such that
 ${\bf p}\subset {\alpha }$, at the same time dividing out a factor $N_\nu/N_B$. 
\begin{eqnarray}
S_+^{(\alpha)}=N_B N_\nu^{-1}\sum_{{\bf p}\subset {\alpha }}b^\dagger({\bf p})a({\bf p})~,~S_-^{(\alpha)}=]S_+^{(\alpha)}]^\dagger
\nonumber\\
S_3^{(\alpha)}=N_B N_\nu^{-1}\sum_{{\bf p}\subset {\alpha }}[b^\dagger({\bf p})b({\bf p})-a^\dagger({\bf p})a({\bf p})]
\nonumber\\
S_0^{(\alpha)}=N_B N_\nu^{-1}\sum_{{\bf p}\subset {\alpha }}[b^\dagger({\bf p})b({\bf p})+a^\dagger({\bf p})a({\bf p})]
\end{eqnarray}
where $a({\bf p})$, $b({\bf p})$ are the annihilation operators for the respective flavor states. We use a similar construction for
anti-particle operators; but at this point we specialize to the case where none are present, for simplicity in sketching out
methods. The anti-particles are easily reinstated for actual applications.
The commutation rules for these collective operators are
\begin{eqnarray}
[S_+^{(\alpha)},S_-^{(\beta)}]=N_B N^{-1}\delta_{\alpha,\beta} S_3^{(\alpha)}
\nonumber\\
\,[S_+^{(\alpha)},S_3^{(\beta)}]=-2 N_B N^{-1}\delta_{\alpha,\beta} S_+^{(\alpha)}
\label{comrs}
\end{eqnarray}
etc.; within each beam, as for the Pauli matrices $\sigma_\pm=(\sigma_1\pm i\sigma_2)/2$ and $\sigma_3$, except for
the factor of $N_B N^{-1}$.
The Hamiltonian in terms of these variables for the case of the
neutral current interactions is then, 
\begin{eqnarray}
H=\sqrt{2} G_F n_\nu N_\nu N_B^{-1} \sum _{\alpha, \beta}\Bigr[  a_{\alpha,\beta} S^{(\alpha)}_+ S^{(\beta)}_-+
 b_{\alpha,\beta} S^{(\alpha)}_3 S^{(\beta)}_3
\nonumber\\
+c_{\alpha,\beta} S^{(\alpha)}_0 S^{(\beta)}_0\Bigr]+H.C.~~~~~~~~~~~~~~~~
\label{genham}
\end{eqnarray}
where $n_\nu$ is the neutrino number density and the matrices $a$ and $b$ are determined by the geometry of subdivisions into beams, taking into account the angular
factors $(1-\cos \theta_{\bf p,q})$. 
In the equations of motion that follow from (\ref{genham}) and (\ref{comrs}), generically
$i dS/dt=[H, S]$, the factor of $N_\nu$ from the Hamiltonian cancels the factor of $N_\nu^{-1}$
from the commutation rules and we obtain, e.g., 
\begin{eqnarray}
i{d \over dt } S_+^{(\beta)}=\sqrt{2}G_F n_\nu N_B^{-1}\sum _\alpha \Bigr [ a_{\alpha,\beta} (S_+^{(\alpha)} S_3^{(\beta)}+
S_3^{(\beta)} S_+^{(\alpha)})
\nonumber\\
-b_{\alpha,\beta} (S_+^{(\beta)} S_3^{(\alpha)}+
S_3^{(\alpha)} S_+^{(\beta)})\Bigr ]\, .~~~~~~~~~~~~~
\label{eom-zero}
\end{eqnarray}
 
The literature cited in refs.\cite{ful1}-\cite{raf4} is based on these equations, but with a further sweeping assumption,
namely that on the right hand side we can replace the expectation values of products of $S$ operators
with the products of the individual expectation values, e. g.,
\begin{eqnarray}
\langle S_+^{(\alpha)} S_-^{(\beta)}\rangle\rightarrow \langle S_+^{(\alpha)} \rangle \langle S_-^{(\beta)}\rangle~,
\label{mean field}
\end{eqnarray}
generally characterized as the mean-field limit.

The diagonal elements of the matrices $a$ and $b$ are zero, due to the $(1-\cos \theta)$ factor in the original
Hamiltonian. Therefore there are no operator-order ambiguities on the right hand side of (\ref{eom-zero}).

Clearly, in the mean field limit when we take pure flavor states as an initial condition we 
have $\langle S^{\alpha}_+(t=0)\rangle=0$.
Then the equations (\ref{eom-zero}) and their coupled counterparts for the operators
$S_3^{\alpha}$ say that nothing whatever happens. But if we took an initial values $\langle S^{\alpha}_+(0)\rangle \ne 0$,
or added to $H$ an ordinary neutrino oscillation term, then, according to the literature 
a bewildering number of things can happen that depend on the interplay of the two parts
of the Hamiltonian. Beginning in sec. 4 we have studied some aspects of these
effects perhaps more systematically than does the previous literature.

\section{3. Scalar particle exchange model}

We take the coupling to a scalar field $\phi$ to the Dirac neutrinos to be $H_I=g \bar \psi \psi \phi$.
Now the effective Hamiltonian for the neutrino-neutrino interaction consists of a contribution from a scalar
particle exchange and one from virtual annihilation.

We shall consider the scalar mass to be of the order of the neutrino energy scale
or less.  The ${\bf p, q}$ dependence in this case is more complex than in the $Z_0$ exchange case.
For a each momentum ${\bf p}$ there are now four states, neutrino and anti-neutrino each with
either normal helicity (- for $\nu$, + for $\bar \nu$) or opposite helicity. For the massless neutrino case, the
coupling to the scalar connects the ``normal" state only to the ``opposite" helicity. However in constructing
the forward Hamiltonian we shall need all 16 of the bilinear products of a creation and an annihilation operator,
 which we designate $\rho_\alpha$. We use a notation in which $a({\bf p})$ annihilates the $\nu$ state with
(normal) negative helicity and $\bar a({\bf p})$ the $\bar \nu$ state of (normal) positive helicity, with $b({\bf p}),
\bar b ({\bf p})$ annihilating the states of opposite helicity. The bilinears, $a^\dagger a, a^\dagger b, \bar a^\dagger b$ etc. have an algebra
\begin{eqnarray}
[\rho_\alpha ({\bf p}),\rho_\beta ({\bf p'})]=\sum_{\gamma =1}^{16} h^{\gamma}_{\alpha,\beta}\rho_\gamma ({\bf p}) \delta_{\bf p,p'}
\label{s-ops}
\end{eqnarray}\, .

The forward effective Hamiltonian for exchange of the scalar particle of mass $m$,  written in terms of these operators, is, 
\begin{eqnarray}
H^{\rm eff}={g \over V} \sum_{\bf p, q} \sum_{\alpha, \beta}\Bigr ({p^\mu q_\mu \over |{\bf p}| |{\bf q}|}\Bigr )\Bigr (\xi^{\alpha,\beta} \rho_\alpha({\bf p})\rho_\beta({\bf q})\Bigr [D_1 ({\bf p , q})\Bigr ]^{-1}
\nonumber\\
+\eta^{\alpha,\beta} \rho_\alpha({\bf p})\rho_\beta({\bf q})\Bigr [D_2 ({\bf p , q})\Bigr ]^{-1}\Bigr )~,~~~~~~~~~~~~~
\label{scalarx}
\end{eqnarray}
where
\begin{eqnarray}
D_1[{\bf p, q}]=({\bf p-q})^2+ m^2-(|{\bf p}|-|{\bf q}|)^2\,,
\label{D1}
\end{eqnarray}
and

\begin{eqnarray}
D_2[{\bf p, q}]=({\bf p+q})^2+ m^2-(|{\bf p}|+|{\bf q}|)^2\,.
\label{D2}
\end{eqnarray}
 
Here the coefficients $\xi$ and $\eta$ are independent of the directions of ${\bf p,q}$, as long as we use helicity states to
define the $\rho_\alpha( {\bf p})$. The $D_1$ term comes from the exchange of the scalar between
any pair of $\nu-\nu$ pair, $\bar \nu-\nu $ pair or $\bar \nu-\bar\nu $ pair.
 \footnote{
Note that the spinor factor in the exchange amplitude is of the form $[\bar u_1({\bf p}) u_2({\bf q}]
[\bar u_3({\bf q}) u_4({\bf p})]$. We use a Fierz transform in reducing to the form shown in (\ref{scalarx}).
We give the rather complex result, for the cases of interest, in sec. 6.}
The $D_2$  term comes from virtual annihilation
of a  $\bar \nu-\nu $ pair.

In contrast to the case of  the $Z_0$ exchange terms of (\ref{ham}) that drove instabilities only in the case of nonistropic
momentum distributions, the form (\ref{scalarx}) will drive instabilities in perfectly isotropic
systems of neutrinos and anti-neutrinos. These instabilities are the main point of the present
paper.

We can then  introduce {\it ab initio} angular averages of the operators in (\ref{scalarx}).
However, also in contrast to the  model of (\ref{ham}), we wish to consider a range of  scalar particle
mass $m$ that ranges from very small, in comparison to the neutrino momenta, to equal or larger.
Thus, while lacking the $\cos_{\bf p,q}$ factor that drives the angular dependences in the former
case there is a continuum of ${\bf |p|,\bf |q|}$ values in the present case; for computational purposes we need to
divide ${\bf p}$ space into a number finite size bins, just as we did the angular space in the simulations
reported in \cite{rfs}. For our primary stability analysis we use only one bin, taking all neutrinos
to be concentrated, say, at the peak of a thermal spectrum ${\bf |p|}=p_0$. We have examined 
simulations of the same problem with several bins and find no qualitative difference in the
results for the instability conditions or results.

In the one-bin simulation the Hamiltonian is

\begin{eqnarray}
H^{\rm eff}=g n_\nu N_\nu \lambda_1\sum_{\alpha,\beta} \xi^{\alpha,\beta} S_\alpha S_\beta
+g n_\nu  N_\nu\lambda_2 \sum_{\alpha,\beta} \eta^{\alpha,\beta} S_\alpha S_\beta \,,
\nonumber\\
\label{approx2}
\end{eqnarray}
where
\begin{eqnarray}
\lambda_{1,2}=\int_{-1}^1 dx_{\bf p,q} D_{1,2}(p_0,p_0,x_{\bf p,q})^{-1}(1-x_{\bf p,q})\,,
\label{lam12}
\end{eqnarray}
and
\begin{eqnarray}
[S_\alpha,S_\beta]=N_\nu^{-1}\sum_\gamma h_{\alpha,\beta}^\gamma S_\gamma \,.
\label{scalarcom}
\end{eqnarray}

\section{4. N neutrino calculations}
In the mean-field (MF) reduction of equations (\ref{mean field}) for the $Z_0$ exchange problem, or the analogous
equations for the scalar exchange case, derived from (\ref{approx2}) and (\ref{scalarcom}), the particle number does not
appear.  We shall discuss solutions in the following sections. But insight is gained by solving the operator equations
for systems with a finite number of neutrinos. This can be done numerically for numbers $N_\nu$
of a few hundred. 
One might have hoped that the rate corrections to the MF approximation 
would be of order $N_\nu^{-1}$ compared to the MF value. The situation turns out to be somewhat more complex,
but the calculations reported below do support the MFA in the domains in which we need it.
Additionally, the finite $N_\nu$ calculations show fascinating connections
between unstable behaviors of the complete system with $\nu$ oscillation terms turned off, and unstable behaviors in the MF approximation when the oscillation terms are reinstated. 

These calculations are based on the observation that the collective operators, $N_\nu S^{(\alpha)}_\pm, 
 N_\nu S^{(\alpha)}_3$ , which obey angular momentum rules, are sums of individual neutrino ``spin"
 operators for the individual modes $\alpha$. There are $2 N_\nu$ basis states, but when the number
 of bins is small,  the size of the problem can be much smaller. To begin with, the total
 ``angular momentum" of each bin, ${ \bf S^{(\alpha)}\cdot  S^{(\alpha)}}$, is conserved. Then if we have
 an initial state for the bin in which every ``spin" is up, this ``angular momentum" has maximum
 value, characterized by quantum number $j_\alpha$. The basis of states for this one bin is thus
 effectively reduced from $2^{N^{(\alpha)}}$ to $N^{(\alpha)}$.  
 
In the very simplest case Friedland and Lunardini \cite{fl} (see also \cite{mck}) found an analytic solution in the limit of a large number of $\nu$'s  for a two-flavor model with no $ \nu$ oscillation terms. They take an initial state with two
unidirectional beams of different flavors, at relative angle $\theta$. The point is to calculate the time scale for the exchange of a macroscopic fraction of the flavors from one beam to the other. In the notation of the last section, we choose variables
and $S_{\pm},S_3,T_\pm,T_3$, where the $S$'s and $T$'s are the collective respective flavor matrices for the two beams.
The operative neutral-current interaction is then the form
\begin{eqnarray}
H= 2^{3/2}G  N (S_+T_-+S_- T_++{a\over 2}  S_3 T_3)\, ,
\label{model0}
\end{eqnarray} 
where we have defined $G= 2^ {-1/2}G_F n_\nu$.
In the standard model case described by (\ref{ham}) the parameter $a$ is unity; this is the case discussed in
ref. \cite{fl}. But the $a<1$ case,
unphysical here, is a simple prototype both for Z exchange models cases with complex angular distributions,
and for all the results for scalar exchange that we shall describe later.  

We take the initial state to be an eigenstate of  $S_3, T_3$ with eigenvalues $S_3=1$ and $T_3=-1$; that is to say,
the respective beams are of pure flavor and of different flavors.  
As displayed in
(\ref{eom-zero}) the equations of motion for these operators, when expresed in terms of neutrino densities, are free of any explicit occurrence of $N_\nu$ (or, equivalently, the volume).
Nonetheless, in the products $S^{(\alpha)}S^{(\beta)}$ on the right hand side of (\ref{eom-zero})  the operators act in the complete $N_\nu$ neutrino space, and the solutions will have $N_\nu$ dependence.
Solving numerically for the case $a=1$, the mixing time is found to be of order,
\begin{eqnarray}
t_{\rm ev}\sim G ^{-1}N_\nu^{1/2}\,,
\label{flrate}
\end{eqnarray}
essentially the result of ref. \cite{fl} achieved by less elegant means.

There can be no physics here, since this mixing time for a system of a very large number of neutrinos
would be longer than that originating in the ordinary cross-section. Thus the truncation to
the forward Hamiltonian, as one might have expected, threw away almost all of the actual flavor exchanges predicted by the full theory. 

In contrast, if $-1<a<1$, the results are very different. The analytic technique of
Ref. \cite{fl} not being applicable here, direct numerical calculation for these cases gives
the mixing time, which we now explicitly define as the time at which $S_3$ passes through the value zero,
as
\begin{eqnarray}
t_1\sim C_a G^{-1} \log N_\nu  
\label{time1}
\end{eqnarray}
The coefficient $C_a$ is of order unity for the fastest case $a=0$. Eq.(\ref{time1}) holds with somewhat smaller coefficient
for $|a|<.9$ in our numerical calculations. Again we are not at the moment suggesting a physical application of (\ref{time1}), 
though logarithms never really get large.

In Fig.1 we show results for values of $N_\nu$=100,400,1600 for the two cases $a=.5$ and $a=1$. The logarithmic 
dependence on $N_\nu$ in (\ref{time1}) is reflected in the equally spaced intercepts of the solid curves, and the $\sqrt{N_\nu}$
dependence in (\ref{flrate}) in the behavior of the dashed curves.
\begin {figure}[ht]
    \begin{center}
       \epsfxsize 2.75in
        \begin{tabular}{rc}
           \vbox{\hbox{
$\displaystyle{ \, { } }$
               \hskip -0.1in \null} 
\vskip 0.2in
} &
            \epsfbox{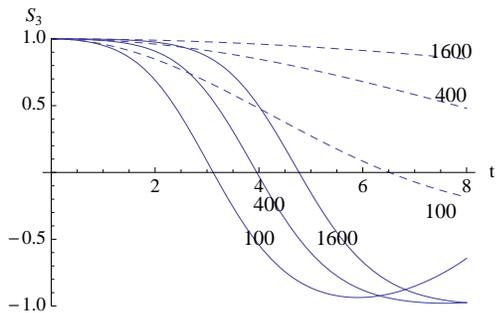} \\
            &
            \hbox{} \\
        \end{tabular}
    \end{center}
\label{fig. 1}
%\vskip 1in
\protect\caption
    {%
Flavor exchange between two beams as a function of time. The curves are labeled
 by the number of neutrinos used in the simulation. The dashed curves are for the case $a=1$ and the
 solid curves are for the case $a=.5$. The time is measured in units $G^{-1}$.
}
\end {figure}

We will not in this section try to analyze how these finite $N$ results could reflect real behavior
of neutrinos that are at large in space. In examples that we develop in the next section we
find that, as one might even expect, when the basic model is augmented with a tiny seed
that can mix the states, the mixing time is given essentially by (\ref{time1}) but with $\log N_\nu$
replaced by $|\log \xi|$ where $\xi$ measures the magnitude of the seed. For the case of $Z_0$ exchange
and flavor mixing, the seeds are the ordinary oscillation parameters. 
For the scalar exchange model 
and helicity reversal, we return later to the question of possible seeds. If nature provided 
none (except for something so super-small that even the logarithm was huge), then possibly
the $\log N_\nu$ model as is would be interesting. We would characterize the situation as
``seeded by quantum fluctuations".

\section{5. Mean field calculations}

Turning to the mean field approximation, 
 we add an ordinary neutrino oscillation term $H_{\rm osc}$ to the Hamiltonian (\ref{model0}), 
\begin{eqnarray}
H_{\rm osc}=\epsilon N_\nu(S_++S_-+T_+/2+T_-/2)
\end{eqnarray}
where we used different oscillation rates for the two different bins, in recognition
of the fact that in applications the oscillation rates depend on energy, and the bins will
typically have different energies.  We use the commutation rules (\ref{comrs}) among the $S$ operators and their
analogues among the $T$ operators to obtain the equations of motion, 
\begin{eqnarray}
i {d \over dt}S_+=G(S_3 T_+ -a T_3 S_+)+ \epsilon S_3~,
\nonumber\\
i {d \over dt}T_+=G(T_3 S_+ -a S_3 T_+)+\epsilon T_3,
\nonumber\\
i {d \over dt}S_3=-2 G(S_-T_+ -S_+T_-)+2 \epsilon(S_+-S_-),
\nonumber\\
i {d \over dt}T_3=-2 G(T_-S_+-T_+S_-)+2 \epsilon(T_+-T_-)\,.
\label{eom}
\end{eqnarray}

With the mean field identities, $S_-=S_+^*$, $T_-=T_+^*$, and the initial conditions
\begin{eqnarray}
S_3(0)=1~,~T_3(0)=-1~,~T_+(0)=T_-(0)=0~,
\end{eqnarray}
the equations (\ref{eom}) 
 determine the evolution of the system. In the absence of oscillations, $\epsilon=0$, none of the variables changes in time for any value of the
 parameter $a$.
 
 In fig. 2 we plot the results of solutions of  the set (\ref{eom}) for the two cases,  for
 three different values of the vacuum oscillation parameter $\epsilon$, with the range of the latter
 chosen to emphasize the similarities between fig. 2 and fig.1.\vspace{.3 in}
\begin {figure}[ht]
    \begin{center}
       \epsfxsize 2.75in
        \begin{tabular}{rc}
           \vbox{\hbox{
$\displaystyle{ \, { } }$
               \hskip -0.1in \null} 
\vskip 0.2in
} &
            \epsfbox{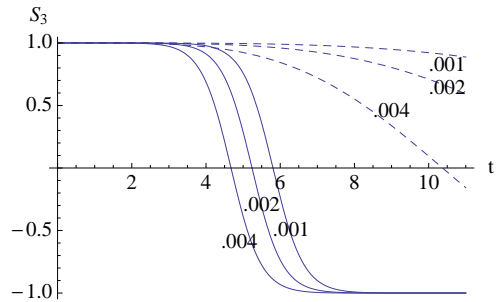} \\
            &
            \hbox{} \\
        \end{tabular}
    \end{center}
\label{fig. 1}
%\vskip 1in
\protect\caption
    {%
As in fig.1, the flavor exchange against time, now from the mean field equations, and labeled
by the strengths of the oscillation parameter $\epsilon /G$, plotted for $a=.5$ (solid), and $a=1$ (dashed)
}
\end {figure} 
Extending the calculation to
 the entire region $10^{-6}<\epsilon<.04$ we find a good fits to the values of the $S_3=0$ intercept
 $t_2^{\rm ev}$  to be given for the $a=.5$ data by
 \begin{eqnarray}
 t_2^{\rm ev}=G^{-1} \log [G/ \epsilon]
 \label{t2}
 \end{eqnarray}
 and for the $a=1$ data by,
 \begin{eqnarray}
 t_3^{\rm ev}=1.85 (G \epsilon)^{-.5}~,
 \label{t3}
 \end{eqnarray}
 
  In spite of the intriguing similarities between fig.1 and fig.2, we still at this point have learned nothing
 about whether or not the mean-field equations give the complete solution
 in the limit $N_\nu\rightarrow \infty$. Ideally, now, we would go back and add the vacuum oscillation   terms to the calculations of the last section and try to replicate the curves of fig. 2. The technical
 problem is the following: for the case with $\epsilon=0$ and with our simple initial conditions
 there were only $N_\nu$ simultaneous coupled ODE's to be solved. When oscillations
 are included, there are $N_\nu^2$ equations. Furthermore, as we see from comparing the $a=.5$ plots
 in figs. 1, and 2, a value $N=100$ is not sufficient to push the transition region with $\epsilon=0$
 out far enough to see the specific effect of $\epsilon$. Therefore our computational limit $N=100$ for the case
with the oscillation term included is insufficient for a
 test of the mean field approach for the values of $\epsilon$ used in fig. 2. 

However, we have been
 able to squeeze in a simulation for a value $\epsilon=.2 G$, in the $N_\nu=100$ case, at times so
 small that the ``finite $N_\nu$" contamination hardly enters (and so short that ordinary oscillation
 for the $G=0$ case is negligible) and we find excellent agreement with the mean field result.
 We have also made a similar check of the MF approximation in a very limited domain in the 
 $a=1$ case. 
 
 The key to the vast qualitative difference in the behaviors for the $a=1$ case and the $a<1$
 case for either of the above approaches is found by doing a standard instability analysis 
 on the equations (\ref{eom}).  Setting $\epsilon=0$
and linearizing in the perturbations $S_\pm,T_\pm$ we see from the last two of the set that, to linear order, 
$S_3$ and $T_3$ remain at their initial values. Then the solutions for $S_\pm,T_\pm$ 
 have simple exponential behavior $e^{\pm \lambda t}$ where 
$\lambda=\sqrt{ (1-a^2)}GN$, and the real exponential for the case $a<1$ signifies an exponential growth in response
to a perturbation,  such as the $\epsilon$ perturbation that gave rise to the rapid mixing in (\ref{t2}). 
The slower mixing in the case $a=1$ of (\ref{t3}), is still much faster than the oscillation time
$\epsilon^{-1}$ in the case that $G/\epsilon=G_F n_\nu / \epsilon>>1$, which holds in the regions of most interest in a supernova. In this case we can roughly say that two zero eigenvalues of the
response matrix of the linearized system are effectively driven complex when $\epsilon$ is
introduced and the system evolves a little, leading to a speed-up but not as dramatic a speed-up.
We characterize this behavior as ``moderately fast", and that of (\ref{t2}) as ``very fast".

In the application to neutrino physics in a supernova, there have been a number of numerical simulations
 \cite{ful1}-\cite{raf4} that produce flavor exchange at a rate much faster than vacuum oscillation rates and occurring somewhat outside the neutrino-surface.  These simulations must divide the neutrino flow into many bins,
since the angular dependences coming from $\cos (\theta_{\bf p,q}) $ in (\ref{ham}) are critical. 
The authors do not analyze their results in terms of the stability matrix of the linearized problem, but it is our conclusion that the closer-in of these phenomena are of our category, ``moderately fast", in the sense that they stem from
a pair of zero eigenvalues. They also appear to fit the $(G_F n_\nu \epsilon)^{-.5}$ behavior
of our simple case.

However, there cases in which the ``very fast" evolution appears not only possible but likely\cite{rfs2}, \cite{rfs}.
This can occur when we have multiple beams simulating an initial flavor dependent angular distribution more disorganized than those assumed in refs. \cite{ful1}-\cite{raf4} (which have perfect cylindrical symmetry).  This appears
to be the case in the
region just below the surface of last scattering for a typical $\nu_e$. In studies reported in 
\cite{rfs} we started with a basic distribution in this region in which
the $\nu_\mu$ and $\nu_\tau$ neutrinos (and anti-neutrinos) have angular distributions biased more outward that than those of $\nu_e$, $\bar\nu_e$. With this distribution, we found ``moderately fast" evolution,
in a number of cases, with the most dramatic results for the ``inverted hierarchy" assumption regarding the neutrino mass matrix. 

Then we added small
random transverse distributions, breaking azimuthal symmetry. About 50\% of the time the behavior shifted to
that of very fast evolution, with mixing lengths much less than 1 cm. These results were robust with respect to introducing a neutrino energy spectrum, through the energy dependence
of the oscillation terms, and the results were essentially the same for the normal and inverted hierarchies. 
In view of the turbulent fluid motion in this region, this suggested complete equalization of the energy
spectra of the various neutrinos emerging from the neutrino-surface, with possible effects on explosion
dynamics and the R process, and with certain effects on the neutrino pulse received on earth.
In contrast to standard MSW transformations, which are
driven completely by the change in background electron-density changes, these instabilities only require
time to pass in order to establish the prerequisites for later sudden changes, with all
model parameters remaining time independent. 
 
 \section{6. Scalar exchange models}

With the results of the previous sections in mind we now investigate the behavior
of the model described in section 3. Beginning from the one-bin equations (\ref{approx2}), (\ref{lam12}) and
(\ref{scalarcom}), we note first that in the mean field interpretation we can, for the purposes of
deriving the equations for the collective variables $S^{(\alpha)}$, cut down notation by eliminating
all sums over modes ${\bf p}$ in their definitions, and at the same time setting $N_\nu=1$ in 
(\ref{approx2}) and (\ref{scalarcom}). 
We give a symbol for each of the
16 bilinear forms in the operators $a$, $b$, $\bar a$, $\bar b$, introduced in sec. 3, and then list all 120 independent relations of the
commutator algebra.  In this notation we have the 16 operators that we divide into three groups,

Group I
\begin{eqnarray}
Q_3=b^\dagger b-a^\dagger a~,~\bar Q_3=\bar b^\dagger \bar b-\bar a^\dagger \bar a\, ,
\nonumber\\
N_0=b^\dagger b+a^\dagger a+\bar b^\dagger \bar b+\bar a^\dagger \bar a\, ,
\nonumber\\
Z=a^\dagger  a + b^\dagger  b-\bar a^\dagger \bar a -\bar b^\dagger \bar b\,.
\label{ops1}
\end{eqnarray}

Group II:
\begin{eqnarray}
Q_+= b^{\dagger}a~,~Q_-= a^{\dagger}b~,~\bar Q_+=\bar b^{\dagger} \bar a~,~\bar Q_-=\bar a^{\dagger}  \bar b
\label{opsII}
\end{eqnarray}

Group III:
\begin{eqnarray}
X_+=\bar b^{\dagger}a~,~X_-= a^{\dagger}\bar b~,~Y_+=b^{\dagger} \bar a~,~Y_-=\bar a^{\dagger}  b
\label{ops3}\,.
\end{eqnarray}

Group IV
\begin{eqnarray}
V_+=\bar a^\dagger a~,~V_-= a^\dagger \bar a~,~W_+= b^\dagger \bar b~,~W_-=\bar b^\dagger b
\label{ops4}\,.
\end{eqnarray}
 
The Hamiltonian (\ref{approx2}) is given in terms of these operators by
\begin{eqnarray}
H^{\rm eff}=g n_\nu \lambda_1 \Bigr[2(X_+Y_++ X_- Y_-) ~~~~~~~~~~~~~~~~~~~~~
\nonumber\\
+(2+\lambda)(V_+W_+ +V_- W_-)+\lambda (X_+X_- +Y_+ Y_-)~~~~~~~~
\nonumber\\
+Q_+Q_++Q_- Q_--Q_3Q_3/2+\bar Q_+ \bar Q_++\bar Q_- \bar Q_-~~~~~~~~~~~~~~~~~
\nonumber\\
-\bar Q_3 \bar Q_3/2+\lambda (Q_+\bar Q_-+Q_- \bar Q_+-Q_3\bar Q_3/2)+N_0^2-\eta Z^2]\,,  ~~~~~~~~~~
\nonumber\\
\,
\label {final ham}
\end{eqnarray}
where $\lambda=\lambda_1/\lambda_2$, $\eta=1/4+\lambda/8$
and,
\begin{eqnarray}
\lambda_1=(2p_0)^{-2}\Bigr [1+{m^2\over 4 {\bf p^2}}\log[{4 {\bf p}^2+m^2\over m^2}] \Bigr]
\nonumber\\
\lambda_2=(2p_0)^{-2}\Bigr [-1+{m^2\over 4 {\bf p^2}}\log[{4 {\bf p}^2-m^2\over m^2}]\Bigr]\,.
\label{lambdas}
\end{eqnarray}
The equations of motion are given as usual by $i d/dt\, S=[S,H]$ where we now use the commutation rules that
come directly from the definitions (\ref{ops1})-(\ref{ops4}).{\footnote{Before making the mean-field approximation, we could worry about the order of factors on the right hand side of the equations.
But the commutator terms that distinguish one operator order from another would give corrections of relative
order $N_\nu^{-1}$ in the end, and thus there is no ambiguity in the mean-field equations.}
 
We shall always suppose that we start at $t=0$ from a state in which all of the operators that mix one helicity with another,
or that mix neutrino with anti-neutrino, are zero.
The reason for dividing the operators into the four groups is that there are two cases in which the problem simplifies a little: a)
when the mean fields of all of the operators of group II and  group IV , namely $Q_\pm, \bar Q_\pm, V_\pm,W_\pm$ remain zero; 
or b) when the operators of group III and group IV  $X_\pm,  Y_\pm, V_\pm,W_\pm$ , remain zero. In either case the
linearized mean-field equations have a growing mode, when we take initial condition of all-active helicities. Both cases 
require a seed to make anything happen. In case b, we believe that a neutrino magnetic moment in a magnetic field can
provide the seed, but discussing it in detail would require backtracking to put angle dependence in the equations
and require binning in angle in the simulations.
We treat case (a) in detail in the next section.

\section{7. The case; $Q_\pm=\bar Q_\pm=V_\pm=W_\pm=0$}
As in the eariier mean-field examples, nothing will happen until we add an analogue of the neutrino
oscillation terms used above. In the present case we use the minimal lepton-number breaking bilinear,
$H'={\epsilon\over m_\nu} \int d{\bf x}\psi ({\bf x})C\psi ({\bf x})$, where $m_\nu$ is the neutrino
mass and $C$ the charge conjugation matrix. This translates, in our notation, into a mixing term in the forward
Hamiltonian,
\begin{eqnarray}
H'=\epsilon  (X_++X_-+Y_++Y_-)
\label{H'}
\end{eqnarray}
This will catalyze reactions in which there is sudden total exchange of helicity between the neutrinos
and anti-neutrinos.  Now defining $G=g n_\nu \lambda_1$, the effective Hamiltonian is,  

\begin{eqnarray}
H=G \Bigr[2 X_+Y_++2 X_- Y_- +\lambda (X_+X_- +Y_+ Y_-)
\nonumber\\
-Q_3Q_3/2-\bar Q_3 \bar Q_3/2 
-\lambda Q_3\bar Q_3/2) -Z^2 \Bigr]  +H'
\,
\nonumber\\
\label {final ham2}
\end{eqnarray}

This, together with the commutation rules, 
\begin{eqnarray}
[X_+,X_-]=(Q_3+\bar Q_3-Z)/2~,
\nonumber\\
\,[Y_+,Y_-]=(Q_3+\bar Q_3+Z)/2~,
\nonumber\\
\,[X_+,Q_3]=[X_+,\bar Q_3]=-X_+~,
\nonumber\\
\,[Y_-,Q_3]=[Y_-,\bar Q_3]=Y_-~,
\nonumber\\
\,[X_+,Z]=2X_+~;~ [Y_-,Z]=2Y_-~,
\label{finalcr}
\end{eqnarray}
determines the equations of evolution.
We easily see that $Q_3-\bar Q_3$ is conserved, and we will always take initial conditions such that
$Q_3(t)=\bar Q_3(t)$. With that substitution (\ref{final ham2})  and (\ref{finalcr}) give,

\begin{eqnarray}
i{d \over dt}X_+= G\Bigr[2Q_3 [ Y_-+(1+\lambda)X_+ ]~~~~~~~~~~~~~~~~~~~~~~~
\nonumber\\
-Z[Y_- +(4 \eta+{\lambda \over 2})X_+]\Bigr ]
+ \epsilon ( Q_3-Z/2)\, ,~~~~~~~~~~~~~~~~
\nonumber\\
\,
\nonumber\\
i{d \over dt}Y_-=G\Bigr [2 Q_3 [ -X_+ -(1+\lambda)Y_-  ]~~~~~~~~~~~~~~~~~~~~~~~~~~
\nonumber\\
-Z[X_++(4 \eta+{\lambda \over 2})Y_-\Bigr]
 - \epsilon  (Q_3+Z/2)\, ,~~~~~~~~~~~~~~~~~~~~~~~~
\nonumber\\
i{d \over dt}Q_3=2G[X_+Y_+-X_-Y_-] ~~~~~~~~~~~~~~~~~~~~~~~~~~~~~~~~~
\nonumber\\
+\epsilon [X_+-X_- +Y_+ -Y_-]\,, ~~~~~~~~~~~~~~~~~~~~~
\nonumber\\
i{d\over dt}Z=\epsilon (-2 X_++2 X_-+2 Y_+-2 Y_-)~.~~~~~~~~~~~~~~~~~~~~
\label{finalem}
\end{eqnarray}
As in the earlier examples, we can investigate the possibility of very fast mixing by looking at the linear perturbations 
away from the trivial solution with $\epsilon=0$, $X_+(t)=Y_-(t)=0$ , and $Q_3(t)=${\it const.} by looking at the
two equations,

\begin{eqnarray}
i {d \over dt} X_+=2G Q_3[Y_++(1+\lambda)X_+],
\nonumber\\
i {d\over dt }Y_-=-2 G Q_3[X_++(1+\lambda)Y_-]~.
\end{eqnarray}

The solutions  have time dependence $\exp [\pm \omega_0 t]$ where $\omega_0=2 G Q_3\sqrt {1-(1+ \lambda)^2}$. If  $-2<\lambda<0$ we thus find 
real exponential behavior that can drive rapid mixing. 
Taking  a value of $\lambda=-1$ appropriate for most of our mass region and solving the full equations (\ref{finalem})
we plot the results in fig.3  for 
a set of values of the ``oscillation" parameter  $\epsilon$, ranging over eight decades. 
\begin {figure}[ht]
    \begin{center}
       \epsfxsize 2.75in
        \begin{tabular}{rc}
           \vbox{\hbox{
$\displaystyle{ \, { } }$
               \hskip -0.1in \null} 
\vskip 0.2in
} &
            \epsfbox{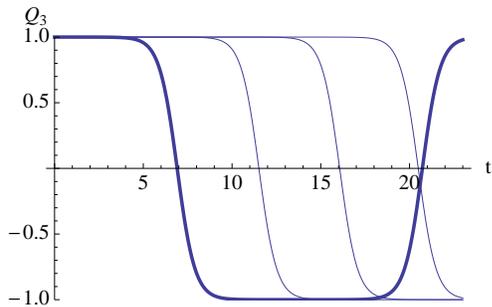} \\
            &
            \hbox{} \\
        \end{tabular}
    \end{center}
\label{fig. 3}
%\vskip 1in
\protect\caption
    {%
Transformation of left handed $\nu$'s and their antiparticles into right-handed counterparts, as
catalyzed by a tiny lepton violating oscillation parameter $\epsilon$. The results for $\epsilon/G=10^{-3}$
are given by the heavy curve. Moving to the right, we show the plots for values, $10^{-5}$, $10^{-7}$ and
$10^{-9}$. Time is measured in units  of $G^{-1}$.
}
\end {figure} 
The intercepts with $Q_3=0$ are well fit by 
\begin{eqnarray}
t_{\rm mix}= \omega_0^{-1} \log[\epsilon/\omega_0]\,.
\label{time}
\end{eqnarray}
Although we have lepton number violation, measured by $Z$, catalyzed by the term (\ref{H'}), at no time in the
evolution does $Z$ become non-negligible, even for our largest value of $\epsilon$.

Finally, we address the fact that the neutrino absolute momenta are to be distributed
in a thermal spectrum, so that we must introduce multiple beams of neutrinos of different values $|{\bf p}|$, exactly as
we introduced multiple angles in the $Z_0$ exchange model \footnote{Our angular averaging was legitimate,
however; note that in the $Z_0$ exchange model we could have taken isotropic
initial distributions and averaged away the $\cos (\theta_{\bf p,q})$ factor in the interaction. But there the interesting
physics is for non-isotropic distributions}. The effective Hamiltonian is now ,

\begin{eqnarray}
H=G N_B^{-1}\sum_{\alpha,\beta}^{N_B}\Bigr (\lambda_1^{\alpha, \beta} \Bigr[2 X^{(\alpha)}_+Y^{(\beta)}_++2 X^{(\alpha)}_- Y^{(\beta)}_- 
\nonumber\\
-Q_3^{(\alpha)}Q_3^{(\beta)}/2-\bar Q_3^{(\alpha)} \bar Q_3^{(\beta)}/2 -Z^{(\alpha)}Z^{(\beta })\Bigr ]
\nonumber\\
+\lambda_2^{\alpha,\beta}\Bigr[ X_+^{(\alpha)}X_-^{(\beta)} +Y_+^{(\alpha)} Y_-^{(\beta)} -Q_3^{(\alpha)}\bar Q_3^{(\beta)}/2 \Bigr ] \Bigr )  
\nonumber\\
+\epsilon \sum_\alpha [(X_+^{(\alpha)}+X_-^{(\alpha)})\pm (Y_+^{(\alpha)}+Y_-^{(\alpha)})\Bigr ]~.
\label {final ham3}
\end{eqnarray}
Here the matrices $\lambda_1,\lambda_2$ are derived from 
$\int d [\cos \theta_{\bf p,q}] D_{1,2}({\bf p,q})$ where the functions $D$ are from (\ref{D1}) and (\ref{D2}), evaluated on a mesh of points $|{\bf p}|_i$, $|{\bf q}|_j$. In the numerical simulation we have taken a subdivision into $N_B$ bins
such that with an initial thermal spectrum every bin is equally occupied. The initial value of each $Q_3^{(i)}$, $\bar Q_3^{(i)}$ is then to be taken as $1/N_B$. If all the matrix elements of $\lambda_1^{i,j}$ are equal to each other and likewise for
$\lambda_2^{i,j}$, then the results for the time evolution of $\sum_i Q_3^{(i)}$ are the same as in the one bin model. But the scatter of values obtained in our procedure can be expected to have some effect.
We have experimented in fitting the parameters $\lambda_1^{i,j}$ for the case of an initial thermal distribution at temperature
T with up to 9 energy bins, each containing the same number of neutrinos, then solving the equations derived from
(\ref{final ham3}) and comparing with the results of the single energy approximation, where we set the parameter
$p_0=2 T$. There is never more than a factor of two difference in the turn-over time, and discrepancies of this
amount are found only in the region $10^{-5}< m/T<10^{-4}$ . We have not attempted the calculation in the
region $m/T<10^{-5}$.

Fig.4 shows an example of a multi-bin simulation in which the elements of the matrix $\lambda_1^{i,j}$ vary over a factor of
6, as compared to a single bin simulation in which $\lambda_1$ is taken as an average. 
\begin {figure}[ht]
    \begin{center}
       \epsfxsize 2.75in
        \begin{tabular}{rc}
           \vbox{\hbox{
$\displaystyle{ \, { } }$
               \hskip -0.1in \null} 
\vskip 0.2in
} &
            \epsfbox{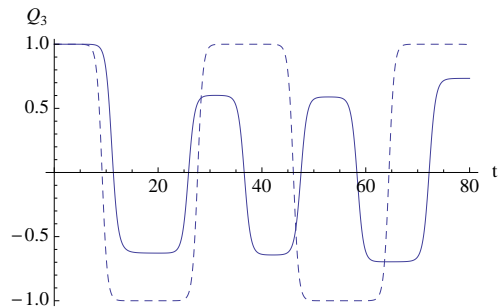} \\
            &
            \hbox{} \\
        \end{tabular}
    \end{center}
\label{fig. 4}
%\vskip 1in
\protect\caption
    {%
The quantities in the same units as in fig. 4. The dashed curve is for the single bin theory
as plotted in fig. 4, here shown for a mixing parameter $\epsilon/G=10^{-4}$. The solid
curve is for the multi-bin model with coupling constants varying over a factor of 6, rather than
for a single $G$. }
\end {figure} 

In all of the situations discussed above, our universe contained only neutrinos and antineutrinos. We believe, however,
that the results are applicable under some conditions to real situations in which there are other particles that interact with neutrinos,
such as the electrons and positrons in the early universe. A condition that we
must demand is that the rate of neutrino reactions from ordinary scattering of neutrinos in the medium
be small compared to the rate of transformation by the coherent mechanism. The limits on the scalar coupling
that we shall quote demands that rate for the coherent process be at least five times 
that based on [(cross-section)$\times$ density].  
However, in the case of the scalar coupling model,
there is a further issue that we must discuss before concluding that we do indeed have a model that predicts
rapid creation of sterile helicity states. When we choose the scalar particle ($\phi$) mass to be substantially less than the
temperature, which we took as 1 MeV in the application, we must worry about the real process $\nu +\bar\nu\rightarrow \phi$;
for equal mixtures of active and sterile particles this reaction rate is exactly of the same order of magnitude as that of our rapid coherent process

In the initial state in our calculation, occupied only by active neutrinos and anti-neutrinos, this process is forbidden,
but as our state with mixed in sterile particles develops, there would be reactions $\nu+\bar \nu_s\rightarrow \phi$
and $\bar \nu+\nu_s\rightarrow \phi$. One would guess that this would not change the qualitative results,
but it would have been good to have an analytic treatment of the effect. The issue, moreover, appears to be essentially
the same as that posed in ordinary neutrino oscillation theory when incoherent scattering is introduced. This has been
discussed many times, going back at least to ref. \cite{tmck}, and we can state a typical outcome as follows:
in the case of a large angle neutrino oscillation where the neutrino scatters as well, with flavor conservation but flavor dependence in
the scattering rate, the scattering makes little difference over a few oscillation periods if the two time scales
are comparable. If the scattering rate is much faster than the oscillation rate than the oscillation rate in the absence
of scattering, then the flavor mixing is slowed, even frozen (in the limit of large scattering). But in our analogous
system, the incoherent process of $\phi$ production is never strong enough to produce a major effect.

\section{8. Discussion}
We have shown that the minimal Yukawa coupling of a Dirac neutrino to a scalar particle can, through a collective
effect, lead to rapid production of 
otherwise nearly sterile right-handed neutrinos, in an initial thermalized state of left-handed neutrinos and anti-neutrinos.
The inverse time scale is essentially of order $(g^2 T/\log[T^{-1}\epsilon]$, where $\epsilon$ measures strength of
lepton-nonconservation from some other source. For any reasonable value of $\epsilon$ other than zero this is
much faster than the rate of production $g^4 T$ in an ordinary
perturbative calculation of $\nu_L+\bar \nu_L\rightarrow \nu_R+\bar \nu_R$ from cross-sections and density.

If we extend the model to include all three neutrino species, universally coupled to the scalar, then we can use the estimate (\ref{time}) to put limits on the parameters $g,\epsilon$ to avoid complete population of the sterile (or ``wrong-helicity")
states, contributing an additional three neutrinos to the mix in the period shortly before decoupling.\footnote{Of course, in the model
we used in this paper the creation of the sterile states depletes the occupancy of the active states. But in the actual 
environment, before $\nu$ decoupling, the ordinary weak interaction rates, although slower than our transformation rate,
will re-populate the active states; in the end we would expect rough thermal occupancy for all states.}  

The estimates do not depend critically
on the mass of the scalar particle as long as it is in the range $0<m<T$. In setting limits on $g$ we demand that the characteristic time for the helicity overturning effect be less than 1/5 
of the collision time for an ordinary reaction, like $e+\nu\rightarrow e+\nu$ in the medium at a temperature of $1\, MeV$.
This gets us to a limit of about $g<10^{-10}$, for values of the lepton number changing parameter $\epsilon>10^{-15} eV$.
This limit is less than those that come directly from particle experiment, and from supernova analysis, the latter at
the $g<10^{-4}$ level. 

Alternatively, if, as seems to be indicated by recent WMAP analyses \cite{wmap}, more that three light neutrino species
are needed, then it would be easy to adjust parameters to get any number between three and six in these
models. 

If couplings of the scalar are taken to change neutrino flavor, then the stability analysis gets very complex
but is still straightforward. 
However the purpose of the present paper was to point out some range of possibilities rather than to try
to do detailed phenomenology, with the secondary purpose to understand better the workings of the much discussed
instabilities in the Z exchange case. In the latter case we have found that the mean-field assumption
is supported by simulations involving a few hundred neutrinos. 

In our detailed model, on which the above estimates are based, the seed was taken as a tiny term that
violates lepton number conservation. It appears that a neutrino magnetic moment in a local magnetic field
could also drive the effect.
Given various more fundamental parameter assumptions, of a kind practiced in the game of ``model building'',
one might be able to argue that one or the other should dominate. A third possibility, briefly discussed at the end of
sec.4 is that quantum fluctuations could serve as a seed.

Other physical systems can show collective behaviors analogous to the ones that we have discussed
for neutrinos. Indeed an exactly analogous case has already been pointed out. Kotkin and Serbo showed
\cite{ks}, in a purely classical
calculation, how a beam of polarized photons can exchange polarizations with another beam of polarized photons,
through the (Heisenberg-Euler) quartic effective \cite{slac} interaction in QED. This conclusion does not require that the beams be monochromatic
or coherent, though in practice the intensities needed for an experiment would surely require laser sources. 
\footnote{The required parameters do not appear to be very far beyond those of the SLAC experiment \cite{slac} 
that detected the reaction $\gamma+\gamma \rightarrow {\rm e}_++{\rm e}_-$, rather misleadingly  reported
as ``photon-photon" scattering . Although few would doubt the correctness or applicability of the
effective 4-$\gamma$ coupling for photons in the ultraviolet, its experimental confirmation would be of some interest.}
As in the case of our very fast models, the calculated rate for the process is
many, many orders of magnitudes faster than the rate for polarization exchange
as calculated from the textbook photon-photon scattering cross-section, by virtue of being lower order 
in the effective coupling constant.
A derivation and generalization of these results following the approach of the current paper can be
found in ref  \cite{rfs4}. This work was supported 
in part by NSF grant PHY-0455918.

\end{document}